\documentclass[11pt]{article}

\usepackage{epsfig,dsfont}
\usepackage{geometry}
\usepackage[T1]{fontenc}
\usepackage[utf8]{inputenc}
\usepackage[english]{babel}
\usepackage{amsmath}
\usepackage{amsfonts}  
\usepackage{epsfig}
\usepackage{calc}
\usepackage{graphicx}
\usepackage{xcolor} 
\usepackage{fancyhdr}
\usepackage{booktabs}
\usepackage{slashed}
\usepackage{bbold}
\usepackage{geometry}
\usepackage{subfig}
\usepackage[titletoc]{appendix}
\usepackage{tikz}
\usepackage{relsize}
\usetikzlibrary{chains, shapes.misc}
\usetikzlibrary{matrix,shapes,arrows,positioning,chains}
\usetikzlibrary{calc}

\begin{document}  
\sffamily

\vspace*{0mm}
 
\begin{center}
{\LARGE
Density of States FFA analysis of SU(3) lattice \vskip2mm
gauge theory at a finite density of color sources}
\vskip15mm
Mario Giuliani, Christof Gattringer
\vskip5mm
Universit\"at Graz, Institut f\"ur Physik, Universit\"atsplatz 5, 8010 Graz, Austria
\end{center}
\vskip10mm

\begin{abstract}
We present a Density of States calculation with the Functional Fit Approach (DoS FFA) in SU(3) lattice gauge theory 
with a finite density of static color sources. The DoS FFA uses a parameterized density of states and determines the parameters
of the density by fitting data from restricted Monte Carlo simulations with an analytically known function. We discuss the 
implementation of DoS FFA and the results for a qualitative picture of the phase diagram in a model which is a further step
towards implementing DoS FFA in full QCD. We determine the curvature $\kappa$ in the $\mu$-$T$ phase diagram and find 
a value close to the results published for full QCD.
\end{abstract}
\vskip8mm

\section{Introductory remarks}

The success of numerical calculations in lattice field theory relies on the availability of probabilistic polynomial algorithms for 
computing observables in a Monte Carlo (MC) simulation. The key point is the interpretation of the Boltzmann factor $e^{-S}$ 
as a probability. However, in some situations the action $S$ acquires an imaginary part that spoils the probabilistic interpretation 
necessary for a MC simulation, a problem usually referred to as ''complex action problem'' or ''sign problem''. 

An important class
of systems with a sign problem are lattice field theories with non-zero chemical potential. In many cases the complex action 
problem is the main obstacle for an ab-initio determination of the full phase diagram at finite density. 
Different methods such as complex Langevin, Lefshetz thimbles, Taylor expansion, fugacity expansion, reweighting, 
and worldline formulations were applied to finite density lattice field theory 
(see, e.g., the reviews \cite{rev1} -- \cite{rev9} at the annual lattice conferences).

Another important general approach are Density of States (DoS) techniques \cite{rev1}, \cite{Gocksch:1987nt} -- \cite{Giuliani:2016tlu}. 
Here we develop further the Density of States Functional Fit Approach (DoS FFA) and apply it to SU(3) lattice gauge 
theory with static color sources (SU(3) LGT-SCS). The DoS FFA was already presented in depth in 
\cite{Mercado:2014dva}  -- \cite{Giuliani:2016tlu} 
and we refer to these papers for a detailed discussion of the method. Here we review 
only the parts specific for the SU(3) LGT-SCS and the respective observables, which are related to the particle number and
its susceptibility  used to determine a qualitative picture of the phase diagram. 

We stress at this point that the results presented here are not meant as a detailed systematic study of the phase diagram 
of SU(3) LGT-SCS, which would imply a controlled thermodynamical limit followed by extrapolating to vanishing lattice spacing.
This paper serves to document the developments and tests of the DoS FFA in a model which 
is a further step towards a Density of States calculation in full lattice QCD at non-zero chemical potential.

\newpage
\section{Definition of the model and the density of states}

We study SU(3) lattice gauge theory with static color charges. The dynamical degrees of freedom are SU(3)-valued gauge links
$\mathrm{U}_\nu(n), \nu = 1,2,3,4$, where $n = (\vec{n},n_4)$ denotes the sites of a $N_s^3 \times N_t$ lattice 
with periodic boundary conditions. 
The action is given by
\begin{equation}
S[\mathrm{U}] = - \frac{\beta}{3} \sum\limits_n \sum\limits_{\mu < \nu} 
\mathrm{Re} \Bigl[ \mathrm{Tr} \, \mathrm{U}_{\mu}(n) \mathrm{U}_{\nu}(n+\hat{\mu})  
\mathrm{U}_{\mu}^{\, \dagger} (n+\hat{\nu})  \mathrm{U}_{\nu}^{\, \dagger} (n) \Bigr] - 
\eta \! \sum\limits_{\vec{n}} \Bigl[ \, e^{ \mu N_t } \, \mathrm{P}(\vec{n}) + \, 
e^{-\mu N_t} \, \mathrm{P}(\vec{n})^\star \Bigr] ,
\label{eq:action}
\end{equation}
where $\beta$ is the inverse gauge coupling, $\eta$ the coupling strength of the static color sources and $\mu$ is 
the chemical potential. The static color sources are represented by Polyakov loops 
$\mathrm{P}(\vec{n})= \frac{1}{3} \mathrm{Tr} \, \prod^{N_t-1}_{n_4=0} \, \mathrm{U}_4( \vec{n} , n_4 )$.
In the action (\ref{eq:action}) the chemical potential $\mu$ gives a different weight to charges $\mathrm{P}(\vec{n})$ 
and to anti-charges $\mathrm{P}(\vec{n})^\star$, such that the theory has  a complex action problem which is equivalent 
to the one of QCD.

For defining the density of states we decompose the action $S[\mathrm{U}]$ into real and imaginary parts and write it in
the form $S[\mathrm{U}] = S_\rho[\mathrm{U}] - i \xi_\mu X[\mathrm{U}]$, where it is easy to see that 
\begin{equation}
\label{eq:action-imaginary}
\begin{split}
S_{\rho}[\mathrm{U}] & \; = \; - \frac{\beta}{3} \sum\limits_n \sum\limits_{\mu < \nu} 
\mathrm{Re} \Bigl[ \mathrm{Tr} \, \mathrm{U}_{\mu}(n) \mathrm{U}_{\nu}(n+\hat{\mu})  
\mathrm{U}_{\mu}^{\, \dagger} (n+\hat{\nu})  \mathrm{U}_{\nu}^{\, \dagger} (n) \Bigr]
- 2\eta \cosh(\mu N_t) \sum\limits_{\vec{n}} \mathrm{Re}[\mathrm{P}(\vec{n})] \; , \\
X[\mathrm{U}] & \; = \; \sum\limits_{\vec{n}} \mathrm{Im}[\mathrm{P}(\vec{n})] \quad \; \; \mbox{and} \; \;
\quad \xi_\mu \; = \;  2\eta \sinh(\mu N_t) \; .
\end{split}
\end{equation}
The functional $X[\mathrm{U}]$ in the imaginary part is bounded, i.e., $X[\mathrm{U}] \in [ -x_{max}, x_{max}]$,
with $x_{max} =  \frac{ \sqrt{3} }{2} N_s^3$. For $x \in [ -x_{max}, x_{max}]$ we 
introduce the weighted density of states $\rho(x)$ as
\begin{equation}
\label{eq:density-definition}
\rho(x) \; = \; \int \! \mathcal{D} [\mathrm{U}] \, e^{-S_{\rho}[\mathrm{U}]} \, \delta(X[\mathrm{U}]-x) \; ,
\end{equation}
where $\mathcal{D} [\mathrm{U}]$ is the product of Haar measures for all link variables.
Exploiting the transformation $\mathrm{U}_{\nu}(n) \rightarrow \mathrm{U}_{\nu}^{\, \star}(n)$ one finds that $\rho(x)$ is an even
function. Thus  the partition sum $Z$ in terms of the density reads
\begin{equation}
Z \; = \; \int \! \mathcal{D} [\mathrm{U}] \, e^{-S[\mathrm{U}]} \; = \; 
\int_{-x_{max}}^{x_{max}} \!\!\!\! dx \; \rho(x) \, e^{i \, \xi_\mu \, x} \; = \; 
2 \int_{0}^{x_{max}} \!\!\!\! dx \, \rho(x) \, \cos(\,\xi_\mu \, x\,) \; ,
\label{partitionsum}
\end{equation} 
and vacuum expectation values of moments of $X[\mathrm{U}] $ can be computed as moments of $x$ 
in the corresponding integrals over 
the density $\rho(x)$. The expression (\ref{partitionsum}) makes clear the emergence of the complex action problem in the DoS 
formulation: The density $\rho(x)$ is integrated with the oscillating function $\cos(\,\xi_\mu \, x\,)$, and from the definition of the 
coupling $\xi_\mu$ in (\ref{eq:action-imaginary}) it is obvious that the frequency of the oscillation increases exponentially with 
$\mu$ (and linearly with $\eta$), such that $\rho(x)$ has to be computed with sufficient accuracy. The recent developments 
of DoS techniques \cite{PhysRevLett.86.2050}  -- \cite{Giuliani:2016tlu} 
are based on new strategies for calculating $\rho(x)$ with very high precision. 

\section{Using DoS FFA for computing the density of states}

For a DoS calculation the density $\rho(x)$ has to be parameterized on the interval $[0,x_{max}]$ in a suitable way. 
For our parameterization we divide the interval $[0,x_{max}]$ into $N$ sub-intervals $I_n \equiv [x_n,x_{n+1}], n = 0,1, \, ... \, N\!-\!1$ 
with $x_0 = 0$ and $x_N = x_{max}$. The density is then written in the form $\rho(x) = e^{-l(x)}$, where $l(x)$ is a 
continuous function that is piecewise linear on the intervals $I_n$. We normalize the density using the condition $\rho(0) = 1$,
which corresponds to 
$l(0) = 0$. Together with the continuity of $l(x)$ this implies that only the slopes $k_n, n = 0,1, \, ... \, N\!-\!1$,
which determine $l(x)$ in the intervals $I_n$ appear as the parameters of $\rho(x)$. A short calculation 
\cite{Mercado:2014dva}  -- \cite{Giuliani:2016tlu} gives the 
explicit form of $\rho(x)$ as function of the $k_n$,
\begin{equation}
\rho(x) \; = \;  A_n \, e^{-k_n x} \quad \mbox{for} \quad x \in I_{n}  \; \quad \; \mbox{where} \quad
A_n \; = \; e^{- \sum\limits_{j=0}^{n-1} \Delta_j (k_j-k_n)} \; .
\label{rhoparameterized}
\end{equation}
Here $\Delta_j \equiv x_{j+1} - x_j$  
denotes the size of the $j$-th interval. We stress that the intervals can be chosen with different sizes such that 
in regions of $x$ where $\rho(x)$ varies quickly a finer discretization can be used.

For computing the slopes $k_n$ in the DoS FFA we use restricted vacuum expectation values 
$\langle \langle X \rangle \rangle_{n}(\lambda)$, $n=0, ... \, N\!-\!1$, 
which depend on a free parameter $\lambda \in \mathds{R}$. They are defined as
\begin{equation}
\label{eq:restricted-value}
\begin{split}
\langle \langle X \rangle \rangle_{n}(\lambda)  = \frac{1}{Z_{n}(\lambda)} 
\int \!  \mathcal{D}[\mathrm{U}] \, e^{-S_{\rho}[\mathrm{U}]+ \lambda \, X [\mathrm{U}] } 
\;  X [\mathrm{U}] \; \theta_n\bigl( X[\mathrm{U}] \bigr) \, , \;
 Z_{n}(\lambda)  = \!\int \! \mathcal{D}[\mathrm{U}] \, 
 e^{-S_{\rho}[\mathrm{U}]+ \lambda \, X[\mathrm{U}] } \, \theta_n\bigl( X[\mathrm{U}] \bigr) ,
\end{split}
\end{equation}
where $\theta_n ( x ) = 1 \text{ for } x \in I_n$, $\theta_n ( x ) = 0 \text{ for } x \not\in I_n$. The free parameter $\lambda$, which
the restricted vacuum expectation values $\langle \langle X \rangle \rangle_{n}(\lambda)$ depend on, 
can be used to probe the system. We stress that 
the restricted vacuum expectation values are free of the complex action problem and can be evaluated with standard 
Monte Carlo calculations.

The key observation of the DoS FFA is that with the parameterization (\ref{rhoparameterized}) the restricted partition sum 
$ Z_n (\lambda)$ and thus the restricted vacuum expectation value 
$\langle \langle X \rangle \rangle_{n}(\lambda) = \partial \ln Z_n (\lambda) / \partial \lambda$ can be computed in closed form. 
It is convenient to shift and rescale the $\langle \langle X \rangle \rangle_{n}(\lambda)$ into a new form $Y_n(\lambda)$ 
for which one finds the explicit expression \cite{Mercado:2014dva}  -- \cite{Giuliani:2016tlu}:
\begin{equation}
Y_n(\lambda) \equiv 
\frac{\langle \langle X 
\rangle \rangle_{n}(\lambda) - \!\!\sum^{n-1}_{j=0} \!\! \Delta_j }{\Delta_n} - \frac{1}{2} =   
\frac{1}{1\!-\!e^{-(\lambda-k_n)\Delta_n}} - \frac{1}{(\lambda\!-\!k_n)\Delta_n} -\frac{1}{2}  \
= h\Big(\!(\lambda-k_n)\Delta_n \!\Big)  , 
\label{eq:sigmoid-function}
\end{equation}
where in the last step we introduced the function $h(s) = 1/(1-e^{-s}) - 1/s - 1/2$. We find that the shifted and rescaled expectation 
value $Y_n(\lambda)$ is given by  $h( (\lambda-k_n)\Delta_n)$, i.e., it depends only on one of the parameters of $\rho(x)$, 
the slope $k_n$ in the respective interval $I_n$. Thus we can compute   
$Y_n(\lambda)$ for several values of $\lambda$ and determine $k_n$ from a fit of the data for 
$Y_n(\lambda)$ with $h( (\lambda-k_n)\Delta_n)$. The function $h(s)$ approaches $\pm 1/2$ for $s \rightarrow \pm \infty$, is 
increasing monotonically and obeys $h(0) = 0$. Thus it is gives rise to simple stable 1-parameter fits and the 
$k_n$ can be determined reliably \cite{Mercado:2014dva}  -- \cite{Giuliani:2016tlu}. 
Analyzing the quality of the fit is an important self consistency check and poor quality of the fit indicates
that the size $\Delta_n$ of the corresponding interval should be chosen smaller \cite{Mercado:2014dva}  -- \cite{Giuliani:2016tlu}. 
Once the slopes $k_n$ are 
computed, the density $\rho(x)$ can be determined with (\ref{rhoparameterized}) and from $\rho(x)$ we can evaluate the observables.

\begin{figure}
\centering
{\includegraphics[width=0.485\textwidth]{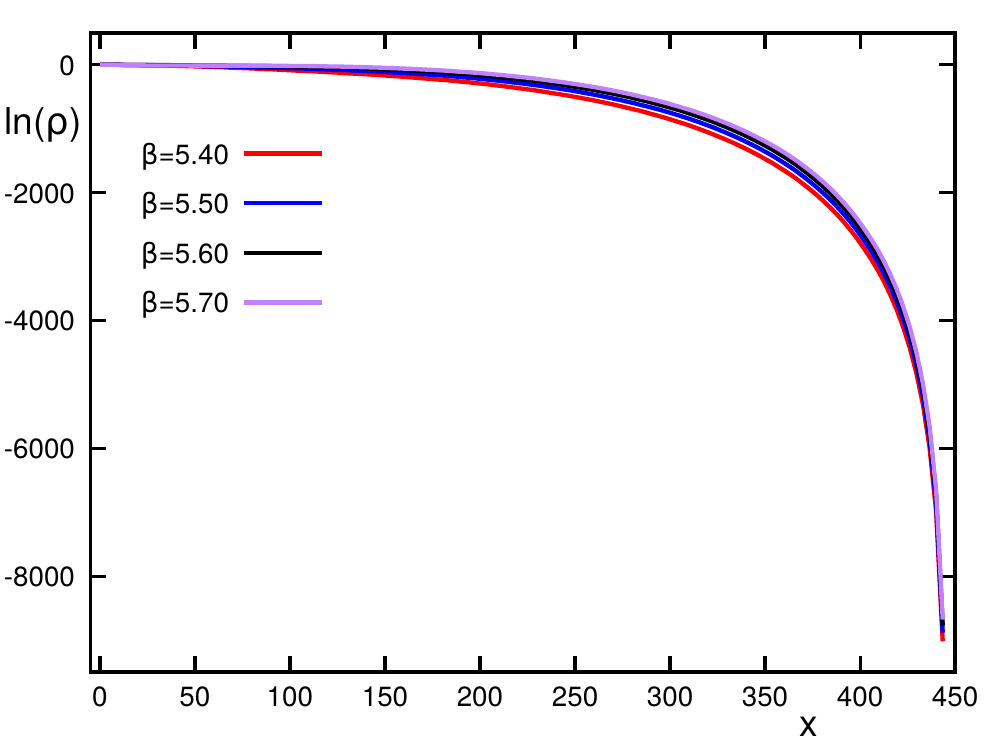}} \,
{\includegraphics[width=0.485\textwidth]{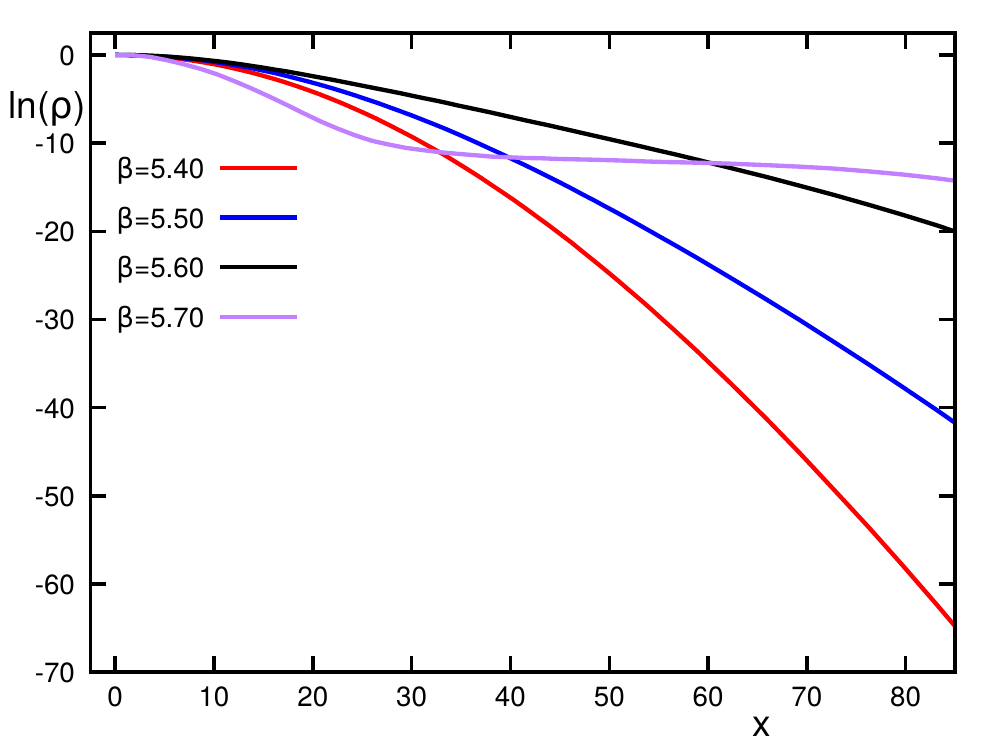}} \,
\caption{Results for the logarithm of the density $\ln \rho(x)$ as a function of $x$ ($8^3\times4$, $\eta=0.04$, $\mu=0.15$ and 
different values of the inverse coupling $\beta$). 
We show $\ln \rho(x)$ for the full range of $x$ in the lhs.\ plot and a zoom into the small-$x$ region 
(rhs.).}
\label{fig:rho}
\end{figure}

Before we come to presenting our results for observables in the next section, we have a look at the density and 
how it changes when varying the parameters. In Fig. \ref{fig:rho} we  show $\ln \rho(x)$ as a function of $x$ for different 
values of the inverse coupling $\beta$ at fixed $\eta = 0.04$, $\mu = 0.15$.  When comparing the different curves for
$\ln \rho(x)$ over the full range of $x$ in the lhs.~plot, the different couplings seem to give rise to essentially the same density. 
However, the zoom into the small-$x$ region (rhs.~plot) reveals that the curve for the largest $\beta$ shows a quite different behavior. 
We will see below that this change corresponds to a phase transition between $\beta=5.60$ and $5.70$. We conclude that inspecting 
the qualitative behavior of the density $\rho(x)$ already reveals physical properties of the system. However, we stress again 
that only the evaluation of physical observables is the true benchmark for a DoS calculation, since the density still has to 
be integrated over with the highly oscillating factors, which tests if the evaluation of $\rho(x)$ is sufficiently accurate.  

We conclude this section with a short comment on the piecewise linear parameterization of the exponent of $\rho(x)$. 
It is clear that the exact result is obtained 
only in the limit where one sends the number of intervals $N$ to infinity and their sizes $\Delta_n$ to 0. 
In our studies for this paper, as well as in \cite{Giuliani:2016tlu}, where we analyzed a related model where we
could systematically compare the DoS FFA results 
to reference data from a dual simulation free of the complex action problem, it was found that the accuracy of the Monte Carlo results 
has a considerably larger effect on the final results than the size of the discretization intervals we use here. More specifically, 
our discretization intervals  $\Delta_n$ were chosen such that an optimal use of the data for $Y_n(\lambda)$ in 
Eq.~(\ref{eq:sigmoid-function}) is obtained: One can show \cite{Giuliani:2016tlu} that the slope of the fit function 
$h((\lambda - k_n)\Delta_n)$ at $\lambda = k_n$ is given by $\Delta_n/12$, and that $\Delta_n/12 \sim 0.1 - 0.5$ 
gives rise to an optimal fit of the data for $Y_n(\lambda)$. This choice from \cite{Giuliani:2016tlu} was implemented 
in our study here.   

\section{Observables and results}

The observables we study are the expectation value of the imaginary part of the Polyakov loop and the corresponding susceptibility. 
Their definitions and expressions in terms of density integrals are given by

\begin{eqnarray}
\hspace*{-4mm}
\langle \mathrm{Im} \, \mathrm{P}  \rangle & \! \equiv \! & - \, \frac{1}{N_s^3} \, \frac{\partial}{\partial \xi_\mu} \, \ln Z \; = \;
\frac{1}{N_s^3} \, \frac{2}{Z} \int\limits_{0}^{x_{max}} \!\!\!  dx \; \rho(x) \, \sin( \xi_\mu \, x ) \; x \; ,
\label{eq:ImP}
\\
\hspace*{-4mm}
{\mathlarger{\mathlarger{\chi}}}_{\mathrm{Im} \, \mathrm{P} } & \! \equiv \! & 
\frac{\partial}{\partial \xi_\mu} \langle \, \mathrm{Im} \; \mathrm{P} \, \rangle \; = \; \frac{1}{N_s^3} \! \left[ \frac{2}{Z} \!\! 
\int\limits_{0}^{x_{max}} \!\!\! dx \, \rho(x) \, \cos(\xi_\mu \, x) \; x^2 + 
\biggl( \frac{2}{Z} \!\! \int\limits_{0}^{x_{max}} \!\!\! dx \, \rho(x) \, \sin( \xi_\mu \, x ) \; x \biggr)^{\!2 \,} \right] \! .
\label{eq:susce_n}
\end{eqnarray}
Note that in leading order we have $\xi_\mu = 2\eta \; \sinh ( \mu N_t) \propto 2 \eta \; \mu N_t$, such that in this order 
$\partial \ln Z / \partial \xi_\mu \propto 1/2\eta \; \partial \ln Z / \partial \mu N_t$, indicating that  
$\langle \mathrm{Im} \, \mathrm{P}  \rangle$ is closely related to the particle number density, and $
{\mathlarger{\mathlarger{\chi}}}_{\mathrm{Im} \, \mathrm{P} }$ to the particle number susceptibility, which 
makes them suitable observables for assessing the phase diagram.

\begin{figure}[t]
\centering
{\includegraphics[width=0.485\textwidth]{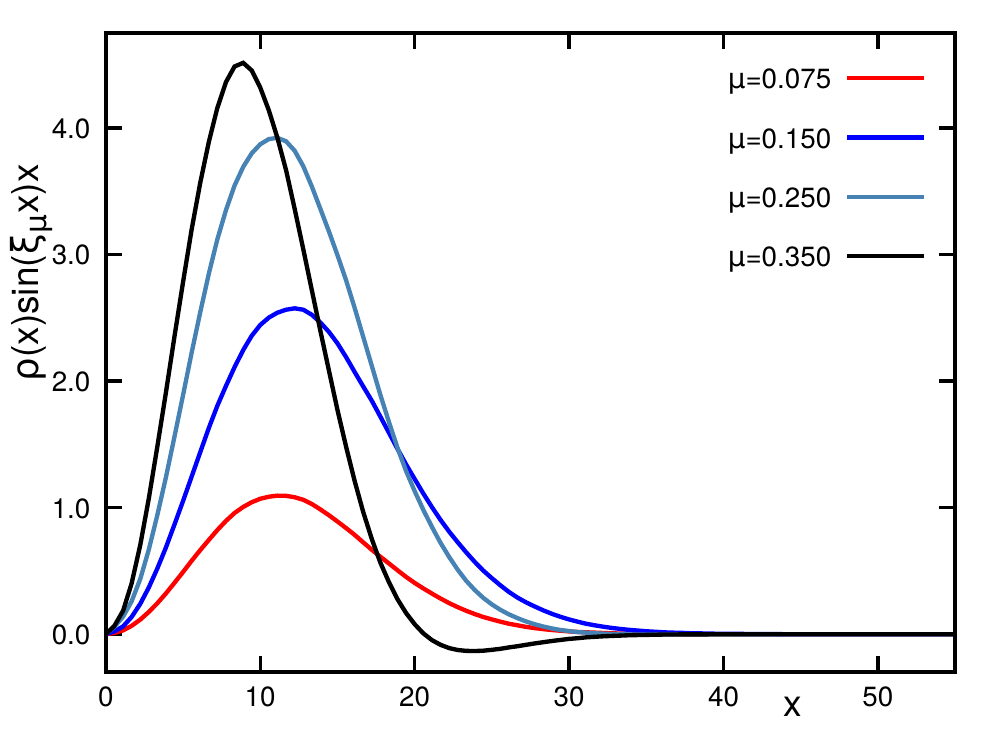}} \,
{\includegraphics[width=0.485\textwidth]{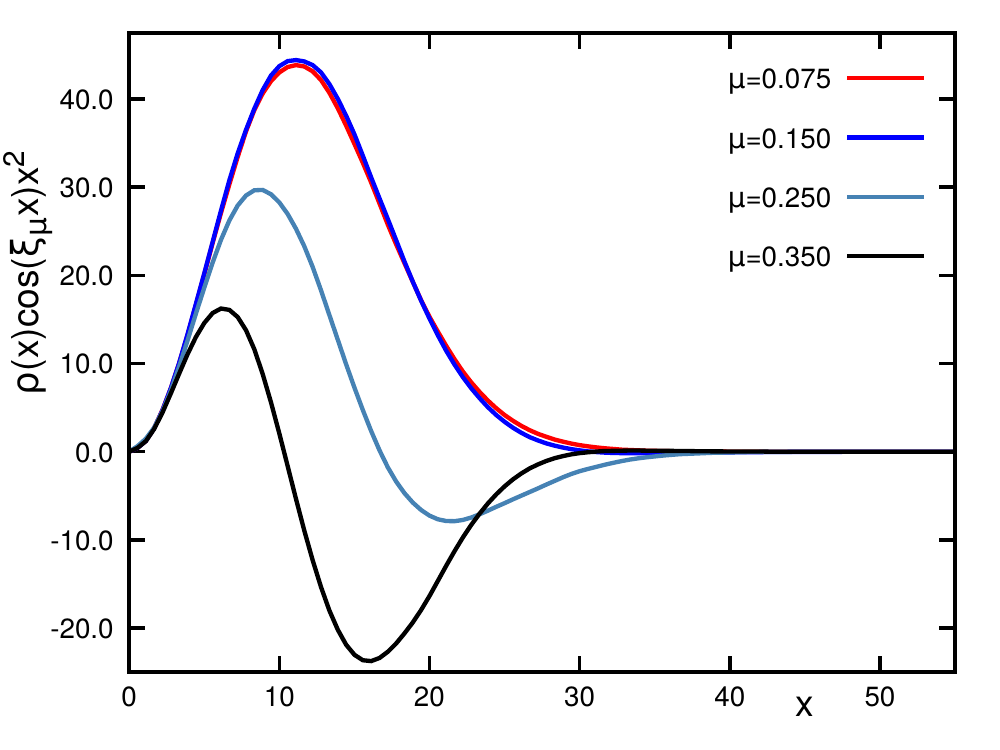}} \,
\caption{The integrand of (\ref{eq:ImP}) $\rho(x) \, \sin( \xi_\mu \, x ) \, x$ (lhs.\ plot), and the integrand 
$\rho(x) \, \cos(\xi_\mu \, x) \, x^2$ of the connected part in (\ref{eq:susce_n}) (rhs.) as a function of $x$ for the different values of 
 $\mu$ used here.}
\label{integrands}
\vskip2mm
\end{figure}

In our numerical simulations we use $N = 256$ intervals for the discretization of $\rho(x)$ for $x \in [0,x_{max}]$ 
(for our $8^3 \times 4$ lattices --
for smaller test volumes see the discussion of the volume dependence in the next paragraph). For each interval we used 
51 values of $\lambda$ and a statistics of 4800 configurations generated with a local restricted Metropolis algorithm. 
The statistical errors for the raw data were computed with a jackknife blocking analysis. For some parameter values 
in the vicinity of the crossover we increased the number of intervals to $N = 384$ and used statistics up to 30000 configurations.

In order to analyze the dependence of the cost on the volume  we implemented 
a small case study at $\beta = 5.45, \mu = 0.25$: In addition to 
$V_4 = 8^3 \times 4$, we also did runs at $V_4 = 6^3 \times 4$ and $V_4 = 4^4$, and adjusted the number of intervals 
$N$ such that the interval size $\Delta_n$ and thus the discretization of $\rho(x)$ remained constant. 
Since $x_{max} = \sqrt{3} N_s^3 / 2$ is proportional to the 3-volume $V_3 \equiv N_s^3$, keeping the discretization of 
$\rho(x)$ constant gives rise to a cost factor proportional $V_3$, which is the cost factor that is specific for the DoS FFA. 
However, as for any other Monte Carlo method, we also need to take into account the longer correlation times and the 
increasing cost of an individual sweep when the volume is increased: Keeping the number of values for $\lambda$ 
fixed at 51, we adjusted the number of Monte Carlo sweeps such that the errors for the density are the same on 
all volumes. This resulted in a statistics of 350, 1500 and 4800 for  $V_4 = 4^4, 6^3 \times 4$ and $V_4 = 8^3 \times 4$, 
which roughly scales like $(V_4)^{1.25}$. Finally, the cost for one local Monte Carlo sweep is proportional to $V_4$, 
such that we expect that the cost of FFA roughly scales like $V_3 \times (V_4)^{2.25}$, where only the factor $V_3$ is specific
for DoS FFA, while the other factor is more general and will also depend on the couplings. 
It is clear that this brief study provides only a very rough 
assessment of the cost and a dedicated analysis is necessary for conclusive cost estimates. More complicated is the 
analysis of the $\mu$-dependence of the cost, since this is expected to very strongly depend on the other parameters. Here we can only 
refer to our study \cite{Giuliani:2016tlu}, where this question could partly be assessed through a comparison with dual simulation data 
in a closely related model. 

\begin{figure}[t!!]
\centering
\includegraphics[width=0.48\textwidth]{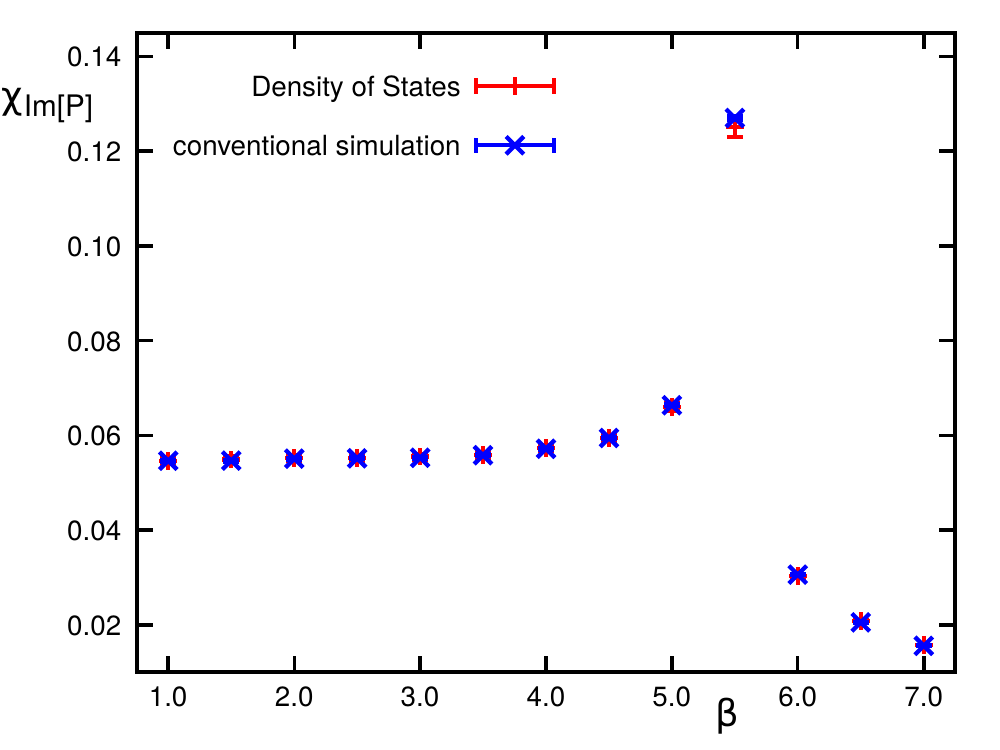}
\hspace{4mm}
\includegraphics[width=0.482\textwidth]{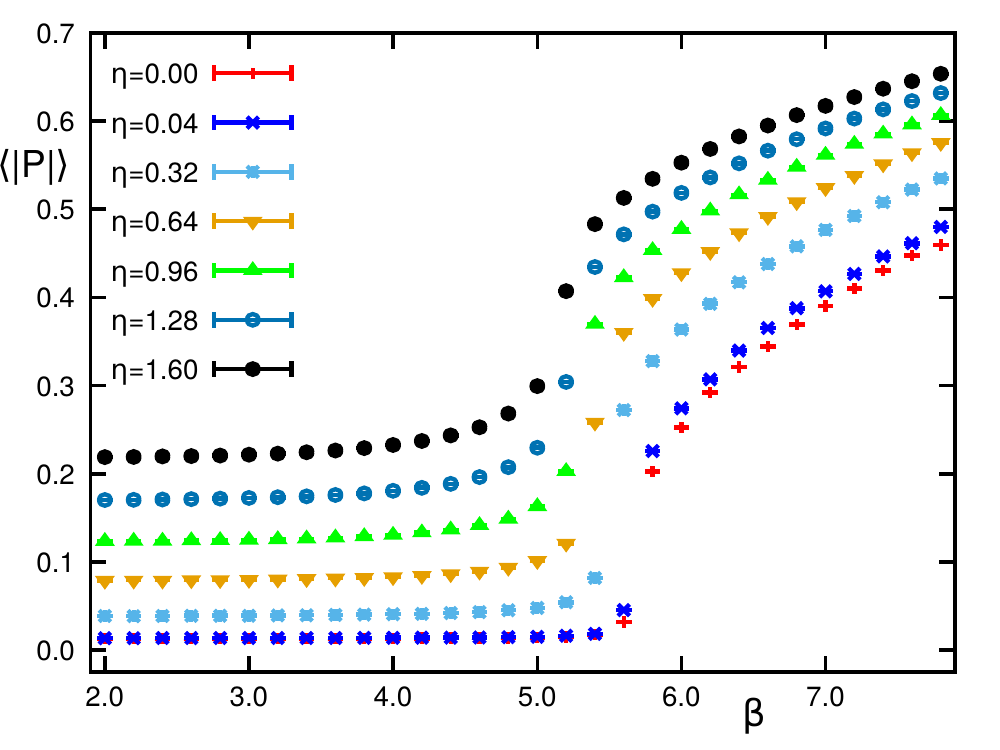}
\caption{Lhs. plot: Comparison of the results for ${\mathlarger{\mathlarger{\chi}}}_{{\mathrm{Im} \; \mathrm{P}}}$ versus $\beta$
computed with the DoS FFA and from a conventional simulation 
for $\mu=0$ ($8^3\times4$, $\eta=0.04$).   Rhs. plot: The absolute value of 
the Polyakov loop versus $\beta$ from a conventional simulation $\mu=0$ ($8^3\times4$, different values of $\eta$).}
\label{fig:conventional}
\end{figure}

Before we come to analyzing the physical observables it 
is interesting to have a look at the integrands of (\ref{eq:ImP}) and (\ref{eq:susce_n}) for the different values of $\mu$ used here. 
In Fig.~\ref{integrands} we show  $\rho(x) \, \sin( \xi_\mu \, x ) \; x$ (lhs.\ plot of  Fig.~\ref{integrands}) and $
\rho(x) \, \cos(\xi_\mu \, x) \; x^2$ (rhs.) as a function of $x$ for different values of $\mu$. While the integrand of (\ref{eq:ImP}) remains
predominantly positive in the range of $\mu$ values we consider here, the integrand of the connected part of (\ref{eq:susce_n}) develops 
essential negative regions illustrating that the complex action problem (sign problem) is challenging for that observable already at the 
values of $\mu$ we access here. When increasing $\mu$ further, both integrands quickly develop a highly oscillating behavior.

From (\ref{eq:ImP}) it is obvious that $\langle \mathrm{Im} \, \mathrm{P} \rangle \equiv 0$ since 
$\xi_\mu = 2\eta \, \sinh ( \mu N_t)$ vanishes at $\mu =0$, while ${\mathlarger{\mathlarger{\chi}}}_{\mathrm{Im} \, \mathrm{P} }$ is 
non-zero also at vanishing $\mu$. Thus we can use ${\mathlarger{\mathlarger{\chi}}}_{\mathrm{Im} \, \mathrm{P} }$ for a first 
assessment of the DoS FFA implementation at $\mu = 0$ where we can compare 
${\mathlarger{\mathlarger{\chi}}}_{\mathrm{Im} \, \mathrm{P} }$ 
to the outcome of a conventional simulation at $\mu = 0$.
The lhs.\ plot of Fig.~\ref{fig:conventional} shows the DoS FFA results 
as well as the results from a standard simulation at $\mu = 0$. We find very 
good agreement of the DoS FFA data with the conventional simulation, which reassures us about the correctness of the 
implementation of DoS FFA and its accuracy, but stress again that the situation at $\mu \neq 0$ is more demanding since there the 
density is integrated with the oscillating factors.

For the runs at chemical potential $\mu \neq 0$ we first have to determine the parameters for the 
simulation, i.e., we need to identify the region with transitory behavior. 
For this purpose we ran a $\mu=0$ conventional simulation at different values of 
$\eta$ to locate a possible transition as a function of $\beta$. Note that with increasing $\beta$ we decrease the lattice spacing
$a(\beta)$, such that increasing $\beta$ corresponds to increasing the temperature $T = 1/(a(\beta) N_t)$.

\begin{figure}[p]
\centering
\includegraphics[width=0.45\textwidth]{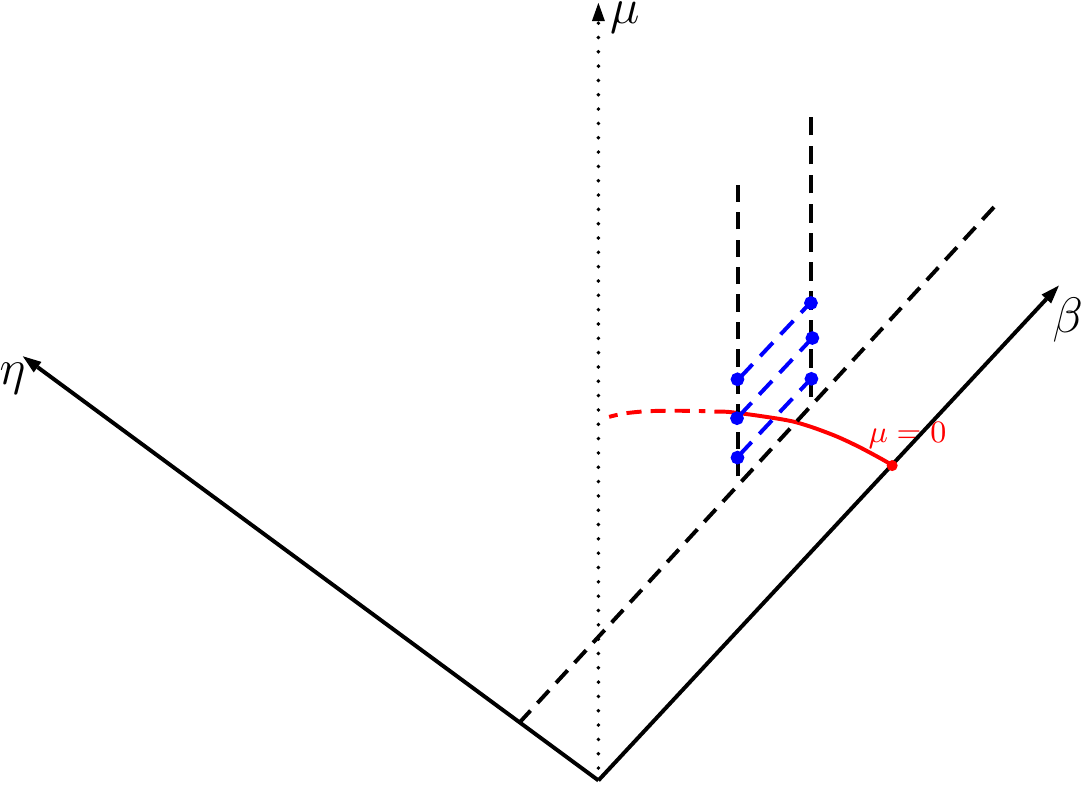}
\hspace{6mm}
\includegraphics[width=0.5\textwidth]{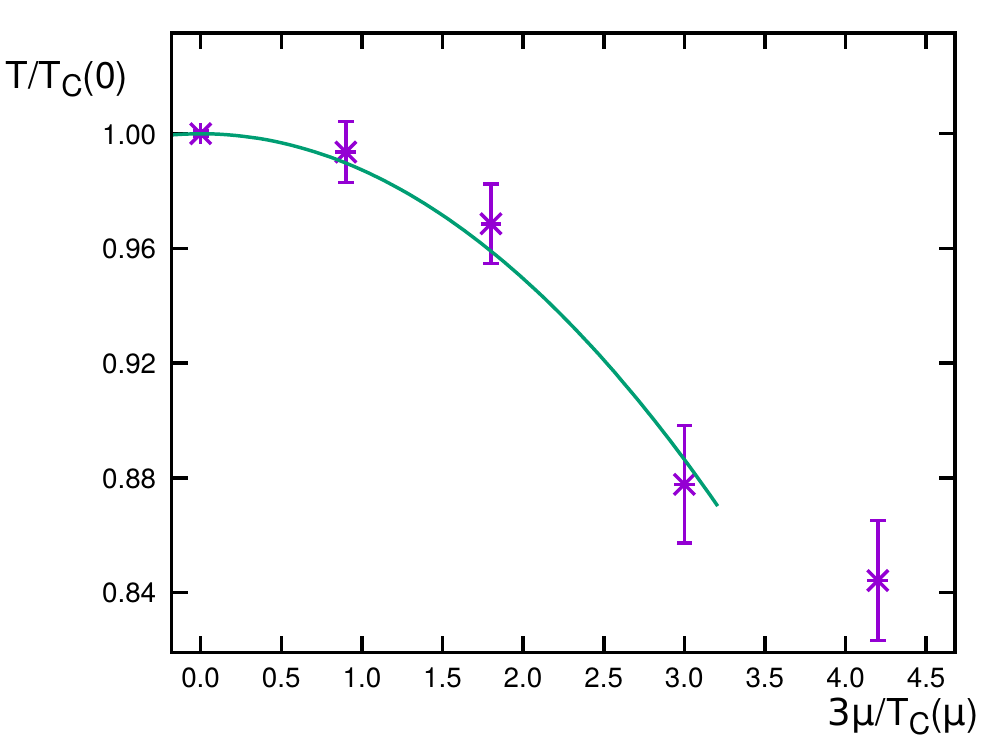}
\vskip5mm
\caption{Lhs.~plot: Schematic sketch of the phase diagram in the $\mu=0$ plane and illustration of the trajectories in coupling space for 
the runs at $\mu > 0$. The red curve in the $\beta$-$\eta$ plane at $\mu = 0$ illustrates the phase boundary between the confined 
(small $\beta$) and the deconfined region. We use a dashed line to indicate  that above some critical $\eta$  one expects only
a crossover type of behavior, while at small $\eta$ the deconfinement transition is of first order (full curve). The blue lines at 
different non-zero values of $\mu$ illustrate the lines in parameter space along which we evaluate observables to probe the 
curvature of the critical surface. Rhs.~plot: Results for the critical line in the $\mu$-$T$ 
plane at fixed $\eta = 0.04$. The data points
on the (pseudo-) critical line were determined as the maxima of a cubic fit of the data points for 
${\mathlarger{\mathlarger{\chi}}}_{{\mathrm{Im} \; \mathrm{P}}}$. In the $\mu$-$T$ plane 
we fit the data with a quadratic polynomial 
to determine the curvature $\kappa$ as explained in the text (the result of this fit is shown as full curve).} 
\label{fig:sketch-phase}
\vskip10mm
\centering
{\includegraphics[width=0.485\textwidth]{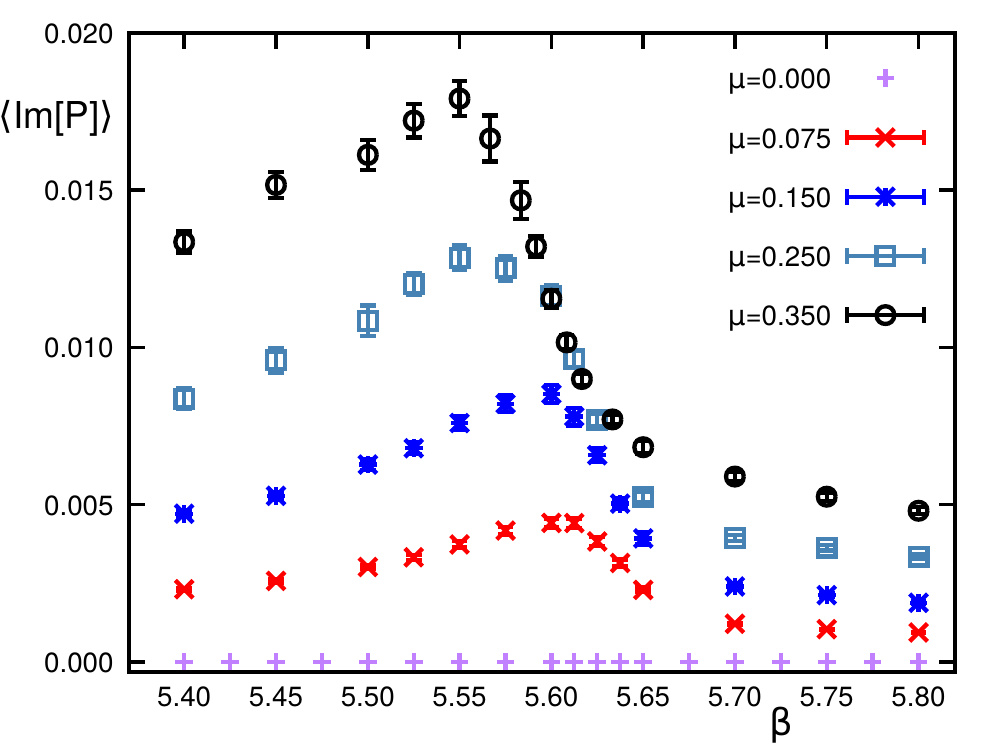}} \,
{\includegraphics[width=0.485\textwidth]{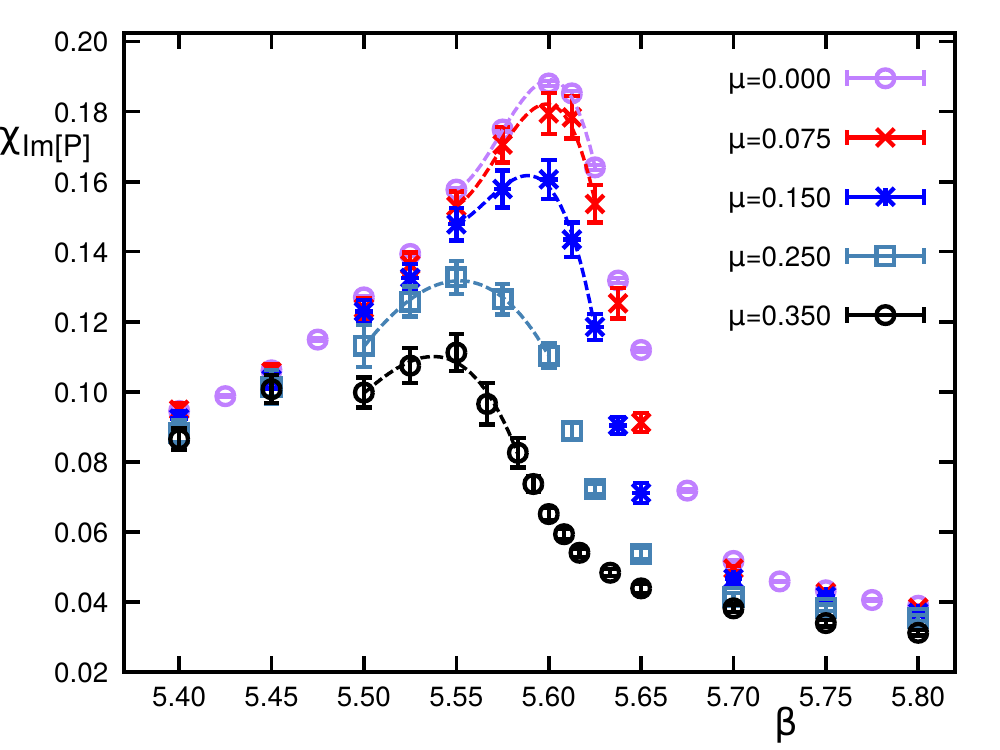}} \,
\caption{Results for the imaginary part of the Polyakov loop
$\langle \mathrm{Im} \, \mathrm{P} \rangle$ (lhs. plot) and its susceptibility
${\mathlarger{\mathlarger{\chi}}}_{\mathrm{Im} \, \mathrm{P} }$ (rhs.) as a function of $\beta$ ($8^3 \times 4, \, \eta=0.04$ and 
several values of $\mu$). The symbols are the results from the DoS FFA calculation and the dashed curves near the maxima of the
${\mathlarger{\mathlarger{\chi}}}_{\mathrm{Im} \, \mathrm{P} }$ represent the fit of the data.} 
\label{fig:Results}
\end{figure}

In the rhs.\ plot of Fig.~\ref{fig:conventional} we show the 
expectation value of the absolute value of the Polyakov loop $\langle | \mathrm{P}| \rangle = 
 \langle | \! \sum_{\vec{n}} \mathrm{P} (\vec{n}) | \rangle/N_s^3$. For all $\eta$ we observe a fast increase of 
$\langle |\mathrm{P}| \rangle$
for values of $\beta$ in the range between $\beta = 5.2$ and 5.8.  For $\eta = 0$ this increase corresponds  to the first 
order phase transition (in the thermodynamical limit) that leads from the confined to the deconfined phase. This transition 
can be understood as spontaneous breaking of the $\mathds{Z}_3$ center symmetry. A non-zero $\eta$ breaks this symmetry
explicitly and thus one expects that above some critical value of $\eta$ the phase transition turns into a crossover.
We also observe that the position of the transition is shifting towards smaller $\beta$ (i.e., towards smaller temperature) 
when increasing $\eta$. We illustrate 
the physical picture in the schematic plot on the lhs.\ of Fig.~\ref{fig:sketch-phase}: The full red curve in the $\mu=0$ plane indicates 
the line of first order transitions, which bends towards smaller $\beta$ when $\eta$ is increased. Based on the argument with the
explicit breaking of center symmetry discussed above, beyond some critical $\eta$ 
we expect only crossover type of behavior which we indicate 
by using a dashed instead of a full curve for that pseudo-critical line. 

In the diagram in the lhs.\ plot of Fig.~\ref{fig:sketch-phase} we also display the $\mu$ axis. When considered in the space of
all three parameters $\beta, \eta$ and $\mu$ we have a (pseudo-) critical surface that separates the confined and 
deconfined phases. This surface runs through the (pseudo-) critical line in the $\beta$-$\eta$ plane. An interesting question 
is how this surface bends for $\mu > 0$, and in the figure we show in blue the trajectories (straight lines parallel to the $\beta$-axis 
at finite $\eta$ and different values of $\mu$)  
along which we compute our observables to explore the bending of the (pseudo-) critical surface.

The results for  $\langle \mathrm{Im} \, \mathrm{P} \rangle$ 
(lhs.\ plot) and ${\mathlarger{\mathlarger{\chi}}}_{\mathrm{Im} \, \mathrm{P} }$ (rhs.) as a 
function of $\beta$ at different values of $\mu$ are presented 
in Fig.~\ref{fig:Results}. Both observables show a maximum at the transition 
between confinement and deconfinement (an exception is $\langle \mathrm{Im} \, \mathrm{P} \rangle$ 
which vanishes at $\mu = 0$ as discussed above). We observe that for increasing chemical potential 
the maxima and thus the transition shift towards smaller $\beta$, i.e., towards smaller temperature. In order to determine the critical line
in the $\mu$-$\beta$ plane we fit the data points near the maxima of ${\mathlarger{\mathlarger{\chi}}}_{\mathrm{Im} \, \mathrm{P} }$
with a cubic polynomial and so determine the peak positions of ${\mathlarger{\mathlarger{\chi}}}_{\mathrm{Im} \, \mathrm{P} }$ 
as a function of $\beta$ for different values of $\mu$, i.e., we determine $\beta_c(\mu)$. 
The results of this determination are used for the plot of the phase diagram in the rhs.~plot of Fig.~\ref{fig:sketch-phase}.

For the presentation of the results for the (pseudo-) critical line in the rhs.~plot of Fig.~\ref{fig:sketch-phase} 
we converted lattice units to physical units using the scale determined for the Wilson gauge action in \cite{scale}. 
On the vertical axis we use the temperature $T$ in units of the critical temperature at vanishing chemical potential, i.e.,
we plot $T/T_c(0)$.
On the horizontal axis we plot the baryon chemical potential $\mu_B = 3 \mu$ in units 
of the critical temperature at that $\mu$, i.e., the combination $ 3\mu/T_c(\mu)$. 
The results of our determination of $T_c(\mu)$ from the maxima of the susceptibility are shown as asterisks in the
 rhs.~plot of Fig.~\ref{fig:sketch-phase}. With increasing $\mu$ we observe the bending of the (pseudo-) critical line towards lower 
temperature values as expected also in full QCD. This bending can be be quantified with the curvature $\kappa$ defined via the 
relation (again we use $\mu_B = 3 \mu$)
\begin{equation}
\frac{T_c(\mu)}{T_c(0)} \; = \; 1 \, - \, \kappa \left( \frac{3 \mu}{T_c(\mu)} \right)^{\!2} \, + \, 
{\cal O} \left( \left( \frac{3 \mu}{T_c(\mu)} \right)^{\!4} \, \right) \; .
\end{equation}
The fit of our data with this quadratic polynomial is shown as the full curve in the rhs.\ plot of Fig.~\ref{fig:sketch-phase}. The 
small-$\mu$ data are described reasonably well and we obtain a value of $\kappa = 0.012(3)$ for the curvature. We stress that this
result is of course a very crude estimate, since we work with only a single lattice spacing and do not attempt a thermodynamic limit. 
Nevertheless it is interesting to note that our result is in the vicinity of the curvature values published for full QCD in different settings,
e.g., $\kappa = 0.0135(2)$ \cite{bonati}, $\kappa = 0.0149(21)$ \cite{WB} and $\kappa = 0.020(4)$ \cite{cea}.

\section{Concluding remarks}

In this letter we have presented an exploratory study where the Density of States Functional Fit Approach DoS FFA 
was implemented for SU(3) lattice gauge theory with static color sources. The purpose is to further develop the DoS FFA 
method towards its use in a full lattice QCD simulation at finite density. The key challenge of DoS calculations is the determination
of the density $\rho(x)$ with sufficient precision, such that one can reliably determine physical observables by integrating $\rho(x)$ 
with the oscillating factor, where the frequency increases exponentially with the chemical potential. 

In our test we demonstrate that for the system of SU(3) lattice gauge theory with static color sources the DoS FFA method can be 
implemented and the accuracy is sufficient for an evaluation of $\langle {\mathrm{Im} \, \mathrm{P} } \rangle$
and ${\mathlarger{\mathlarger{\chi}}}_{\mathrm{Im} \, \mathrm{P} }$. Comparing the DoS results to a conventional simulation at 
$\mu = 0$ shows good agreement and for $\mu >0$ the observables and the critical line could be determined up to moderately large
values of $\mu$. A determination of the curvature $\kappa$ in the $\mu$-$T$ phase diagram gives a value which is surprisingly 
close to the results published for full QCD.

We stress again, that the results presented here should not be considered as a final determination of the phase structure of SU(3) 
lattice gauge theory with static color sources, since infinite volume extrapolation and continuum limit were not attempted here. The
purpose of the paper is to document the further development of DoS FFA and to explore the steps necessary towards getting the 
method ready for a full QCD calculation.

\vskip5mm
\noindent {\bf Acknowledgments:}  This work was supported by the Austrian Science Fund, 
FWF, DK {\it Hadrons in Vacuum, Nuclei, and Stars} (FWF DK W1203-N16) and we thank 
Kurt Langfeld, Biagio Lucini and Pascal T\"orek for discussions.

\end{document}